# Size Effect on Order-Disorder Transition Kinetics of FePt Nanoparticles

Shuaidi Zhang[1],   Weihong Qi[1, 2, 3]*,   Baiyun Huang[2],

[1] School of Materials Science and Engineering, Central South University, Changsha, 410083, P. R. China

[2] State Key Laboratory of Powder Metallurgy, Central South University, Changsha 410083, P. R. China

[3] Key Laboratory of Non-ferrous Materials Science and Engineering, Ministry of Education, Changsha, 410083, P. R. China

\* Corresponding author:   qiwh216@csu.edu.cn    (W. H. Qi)

**Abstract:**

The kinetics of order-disorder transition of FePt nanoparticles during high temperature annealing is theoretically investigated. A model is developed to address the influence of large surface to volume ratio of nanoparticles on both the thermodynamic and kinetic aspect of the ordering process; specifically, the nucleation and growth of $L1_0$ ordered domain within disordered nanoparticles. The size- and shape-dependence of transition kinetics are quantitatively addressed by a revised Johnson-Mehl-Avrami equation that included corrections for deviations caused by the domination of surface nucleation in nanoscale systems and the non-negligible size of the ordered nuclei. Calculation results based on the model suggested that smaller nanoparticles are kinetically more active but thermodynamically less transformable. The major obstacle in obtaining completely ordered nanoparticles is the elimination of antiphase boundaries. The results also quantitatively confirmed the existence of a size-limit in ordering, beyond which, inducing order-disorder transitions through annealing is impossible. A good agreement is observed between theory, experiment and computer simulation results.

**Keywords**: nanoparticles, iron-platinum alloy, order-disorder transition, size-dependence, kinetics

## I. Introduction

Ferromagnetic FePt bimetallic nanoalloys have important applications in future ultrahigh-density magnetic storage devices.[1-6] Spin-orbit coupling of Fe and Pt atoms in chemically ordered $L1_0$-FePt phase[7] gives rise to the highest uniaxial magnetocrystalline anisotropy (MCA) energy ($10^6$ to $10^7$ J/m$^3$) currently known,[8-10] allowing the alloy to remain magnetically stable under room temperature even when reduced to extremely small sizes (~3.5 nm in diameter).[11-15] Theoretically, successful utilization of $L1_0$ ordered FePt nanoparticle array will enable data storage density exceeding 1 T bit/in$^2$.[16] Motivated by the drastically superior performance FePt nanoparticles promised, intense researches have been directed at their chemical preparation.[17-19] Various methods including solution phase decomposition/reduction and salt co-reduction have been continuously perfected, yielding increasingly better size, composition and dispersity control.[20-24] Yet unfortunately, as-synthesized FePt particles are in the disordered A1 form, which is magnetically soft. High temperature annealing is thus applied to induce order-disorder transition in the nanoparticles so the desired $L1_0$ structure can be obtained.[25] Despite prevalent use of the technique, exact kinetic process governing the order-disorder transitions during annealing remains poorly understood. The surface tension and limited size of nanoparticles make it difficult to apply traditional kinetic theories of solid-state phase transitions. Meanwhile, suppression of the equilibrium order-disorder transition temperature in nanoscale FePt systems is widely reported.[26, 27] Both coordination deficiency of surface atoms and large surface to volume ratio in nanoparticles are thought to have destabilizing effects on the $L1_0$ structure, suggesting order-disorder transitions in nanoscale alloy systems may deviate significantly from those in bulk materials. Even though several theoretical works have made considerable progress in the prediction of thermodynamic stability of ordered nanoparticles,[28-30] to our best knowledge, there is currently no systematic theory capable of providing quantitative descriptions for the size-dependent transition process.

In this article, we propose a theoretical kinetic model that is specifically developed to account for the influence of size-dependence, surface effects and geometrical limitation of nanoparticle systems on their order-disorder transitions. The model is capable of quantitatively addressing the nucleation and domain growth within nanoparticles during their transformation, and is highly compatible with recent Monte-Carlo simulations.[31, 32]. By clarifying the transition kinetics, the model also offered plausible explanations to current experimental observations such as the existence of multiple ordered domains in a single nanoparticle,[33, 34] partial ordering[35] and ordering size limits.[35]

**II. Methods**

The thermodynamic driving force of the order-disorder transition can be expressed as Gibbs free energy difference between the two phases △$G_v$ = $G_{disorder}$ − $G_{order}$. The entropy portion of Gibbs free energy is comprised of three parts, configuration entropy $S_{conf}$, vibration entropy $S_{vib}$ and electron entropy $S_{ele}$, i.e., $S_{total}$= $S_{conf}$ + $S_{vib}$ + $S_{ele}$. Comparing to the other two, the influences of $S_{ele}$ is negligible. Furthermore, because order-disorder transition includes only permutation changes under the same temperature, the vibrational entropies of the two phases effectively cancel each other out. According to the definition, $S_{conf}$=$k_B$lnΩ, with Ω representing the total number of possible lattice arrangements and $k_B$ representing the Boltzmann constant. Therefore, for a completely disordered particle constituted of N atoms, $S_{conf}$=$Nk_B$ln2, and for a completely ordered particle, $S_{conf}$=0. [27, 36] Thus, the Gibbs free energy difference per unit volume for such a spherical nanoparticle can be written as △$G_v$ = 3(N△H-$TNk_B$ln2)/4π$R^3$, where R is the radius of the nanoparticle and △H is the unit enthalpy difference between the two phases. At Transition temperature $T_c$, △$G_v$ = 0 and △H=$T_c k_B$ln2. The temperature dependence of △H has trivial effects here, so △H is regarded as a constant. If we approximate N as $R^3/r_0^3$ ($r_0$ is the averaged atomic radius), we get:

$$\Delta G_v = \frac{3k_B ln2 \cdot (T_c - T)}{4\pi r_0^3} \tag{1}$$

For an ordered domain to successfully nucleate within a uniform solid medium, the Gibbs free energy reduced by the presence of the ordered nuclei needs to outweigh the Gibbs free energy gained by both the creation of $L1_0$-fcc interfaces and the introduction of non-compatible lattices[37]:

$$\Delta G^f = \frac{-4\pi r^3}{3}\Delta G_v + 4\pi r^2 \gamma + \Delta G_e \qquad (2)$$

In Eq (2) the formation Gibbs free energy of the nuclei is denoted as $\Delta G^f$, nuclei radius as r, the unit area Gibbs free energy of $L1_0$-A1 interface as γ, and the elastic Gibbs free energy as $\Delta G_e$. Based on our calculation (see supplemental materials), at typical nucleation radius (~1 nm), $\Delta G_e$ is less than 2% of the interface free energy γ, and can be safely dismissed.[37-41] Therefore, Eq (2) becomes:

$$\Delta G^f \approx \frac{-4\pi r^3}{3}\Delta G_v + 4\pi r^2 \gamma \qquad (3)$$

Solving the equation $\frac{\partial}{\partial r}\Delta G^f = 0$ yields the critical nuclei radius $r_c$ = 2γ/$\Delta G_v$. Incorporate the solution and Eq (1) into Eq (3), we get the critical nuclei formation energy:

$$\Delta G^{f,c} = \frac{16\pi\gamma^3}{3(\Delta G_v)^2} = \frac{256\pi^3 r_0^6 \gamma^3}{3(3k_B ln2 \cdot (T_c - T))^2} \qquad (4)$$

From Classic nucleation theories,[37] we know the surface nuclei formation energy can be written as:

$$\Delta G^f_{surface} = \left(\frac{-4\pi r^3}{3}\Delta G_v + 4\pi r^2 \gamma\right)\frac{2 - 3\cos\theta + \cos^3\theta}{4} \qquad (5)$$

where θ represents the wetting angle. A balanced surface tension requires cosθ = ($\gamma_d$-$\gamma_o$)/$\gamma_{int}$. $\gamma_d$, $\gamma_o$, $\gamma_{int}$ denote the surface energies of disordered free-surface, ordered free-surface and order-disorder interface respectively. Previous works[42] have shown free-surface energy is largely dependent on the stoichiometry of the nanoparticles and the coordination number of surface atoms. Since $L1_0$-A1 transition induces trivial morphological changes, and recent experimental results[35] show there are no surface segregations in equiatomic FePt nanoparticles, we regard the energy difference between the two kind of free-surfaces as negligible comparing to the energy of order-disorder interfaces, and have θ =π/2 as an approximation. With this notion, it is easy to see that the critical nuclei formation energy on the surface would equal $\Delta G^{f,c}$/2.

It is already established[37] that critical nuclei formation energy $\Delta G^{f,c}$ relates to nucleation rate "I" in the following fashion:

$$I = \omega exp(-\frac{\Delta G^M}{k_B T}) \cdot C exp(-\frac{\Delta G^{f,c}}{k_B T}) \qquad (6)$$

The term $\omega exp(-\frac{\Delta G^M}{k_B T})$ is a diffusion related term, and C is the number of nucleation site available. For homogeneous nucleation at the core, it is the number of atoms per unit volume; while for heterogeneous nucleation at the surface, it is the number of surface atoms per unit volume. The diffusion term remains the same. By considering the difference in critical nuclei formation energy and approximating C by $\frac{3}{R}(\pi r_0^2)^{-1}$ and $(\frac{4\pi r_0^3}{3})^{-1}$ respectively for surface and homogeneous nucleation in spherical nanoparticles, we can write:

$$I_{surface} = exp\left(\frac{\Delta G^{f,c}}{2k_B T}\right) \cdot \frac{4r_0}{R} \cdot I_h \qquad (7)$$

This means nucleation rate at the surface ($I_{surface}$) is exponentially higher than the homogeneous nucleation rate ($I_h$). In fact, the prioritization of heterogeneous nucleation has long been demonstrated in classic theories[37]. At the degree of supercooling where heterogeneous nucleation begins, homogeneous nucleation is not even detectable. In bulk materials, the heterogeneous nucleation is facilitated at grain boundaries, free surface only accounts for a negligible proportion of the nucleation sites. However, nanoscale systems, especially nanoparticles, are usually defect-free[43] and have a very large surface to volume ratio compared to bulk systems. So, it is important to note that for nanoscale systems, phase change nucleation sites are almost entirely provided by free surfaces.

To better illustrate the size-dependence of surface nucleation, we hereby introduce the area specific notion of surface nucleation rate, "$I_s$", defined as number of nuclei per unit area per unit time. Therefore, $I_{surface}$ = (surface area/volume)$I_s$ and for nanoparticles, $I_{surface}$ = (3/R)$I_s$.

Another factor at play is the suppression of transition temperature "$T_c$" in nanosystems. The effect is ubiquitously observed in nanoparticles, suggesting their large surface to volume ratio and enhanced surface effects might thermodynamically destabilize the ordered phases. Based on the Generalized-Bond-Energy (GBE) model, Qi and coworkers[36] proposed a method to theoretically calculate the size dependence of the transition temperature. By applying their method to the FePt

system, we obtained:

$$T_c = (1 - \frac{1.691 r_0}{R})T_{c,b} \quad (8)$$

$T_{c,b}$ is the transition temperature for the bulk material. Apparently, when the particle radius R approaches the magnitude of atomic radius $r_0$, as in nanoparticles, $T_c$ becomes significantly correlated with R.

The Johnson-Mehl-Avrami(JMA) Equation[44] is often used in literature to describe first-order phase transformations. However, the original equation ignores both surface nucleation and the initial size of the nuclei, which play vital roles in the transformation of nanoparticles. From Eq (1), we could see the critical radius of a stable nucleus can be expressed as:

$$r_c = \frac{8\gamma \pi r_0^3}{3k_B ln2 \cdot (T_c - T)} \quad (9)$$

Based on our calculation, this value is usually around 1 nm. Yet many nanoparticles themselves are only a few nanometers across. This suggests in nanoscale systems, the nucleation process alone completes the majority of the transformation. While further expansion of the nuclei serves only to fill the finite gaps separating each nucleus.

To revise the JMA equation for its application in nanoscale systems, the radius of the as-grown nuclei in the original equation should be formulated as g(t-τ)+$r_c$ instead of g(t-τ), where t is the system time, τ is the time of nuclei formation and g is the domain expansion speed. Consequently, if we ignore domain overlapping, the volume fraction transformed within this time interval, $X_e$, can be differentially expressed as:

$$dX_e = \frac{2\pi}{3}[g(t-\tau) + r_c]^3(\frac{3}{R}I_s)d\tau \quad (10)$$

Note that the growth of surface nucleated domains are modeled as semi-spheres rather than complete spheres. Integrate Eq (10) from 0 to t, we obtain:

$$X_e = \frac{\pi}{2gR}I_s[(gt + r_c)^4 - r_c^4] \quad (11)$$

However, the growth of the ordered domains can only proceed into the untransformed volume. To tackle this issue, the original JMA equation approximated the transformation rate to be directly proportional to the untransformed volume fraction $dX = (1-X)dX_e$, where X is the net transformed volume fraction. Solve this equation and the JMA equation can be rewritten as

$$-\ln(1-X) = \frac{\pi}{2gR} I_s \left[ \left( gt + \frac{8\gamma\pi r_0^3}{3k_B \ln 2 \cdot (T_c - T)} \right)^4 - \left( \frac{8\gamma\pi r_0^3}{3k_B \ln 2 \cdot (T_c - T)} \right)^4 \right] \quad (12)$$

The size effect on the kinetic ordering process is well illustrated in Eq (12), when annealing temperature and time are fixed, the transformed volume fraction X becomes a strict function of R (Since $T_c$ is a function of R, and the rest of the parameters only depend on temperature), meaning particles with different sizes respond to annealing differently.

For defect-free single crystal bulk systems capable of homogenous nucleation, we can ignore the initial size of the nuclei "$r_c$", replace surface nucleation "$(3/R)I_s$" with a general homogeneous nucleation rate "$I_h$" and model the growth of nuclei as spheres rather than semi-spheres. Eq (12) then reverts to the Johnson–Mehl equation: $X=1-\exp[(-\pi/3)I_h g^3 t^4]$, which is the most basic form of JMA equation for describing homogeneous nucleation in macroscopic defect free systems(see supplemental materials & Figure S1).

**III. Results and Discussion**

For FePt nanoparticles, $r_0$=0.1337 nm,[45] and the bulk order-disorder transition temperature $T_{c,b}$ is at 1570 K. Muller *et al.* have developed an FePt lattice energy model[31] that enables the calculation of antiphase boundary (APB) energies. In their model, two types of non-conservative ABP energies are expressed as $4J_1 - 4J_2 \pm 2/3h$ per square lattice constant $a_0^2$. $J_1$ and $J_2$ are determined to be 0.0465 eV and −0.0093 eV respectively. In our model, the term γ denotes the order-disorder interface energy. On the disordered side of the interface, the lattice sites are occupied by Fe or Pt atoms with an equal probability. So we take the averaged value of the energies of the two types of APBs as the value of γ. Thus, $\gamma=4J_1 - 4J_2=1.58$ eV/nm$^2$. To correlate with the unit used, we take the value of Boltzmann's constant to be 8.6173×10$^{-5}$ eV/ K.

As for surface nucleation rate, Li *et al.* reported[46] the temperature dependence of nucleation rate in 20nm thick FePt nanofilms. The nucleation rate $I_{film}$ is directly measured by applying X-ray diffraction, and followed the Arrhenius law $I_{film} = I_0\exp[E_n/(-k_BT)]$, with an activation energy $E_n$ of 0.5 eV and a pre-exponential factor $I_0$ of 0.1618. As previously noted, in nanoscale systems, surface nucleation dominates. So $I_{film}$ gives direct information about $I_s$. Therefore, for nanoparticles

$$I_{total} = \frac{3}{R}I_s = \frac{3}{R}\frac{20}{2}I_{film} = \frac{4.854}{R}\exp\left(\frac{0.5}{-k_BT}\right) \tag{13}$$

The size and temperature dependence of the total nucleation rate "$I_{total}$" is plotted in Figure 1. Apparently, the rate increases with the decrease of particle size and the increase of temperature. Also, it is much more responsive at high temperatures or in particles with small radii. All these characteristics arise from the expansion of surface to volume ratio as particles get smaller and the Boltzmann factor relationship between temperature and nucleation.

By combining Eq (9) and Eq (8), we calculated the critical nuclei radius for different particle size R and different annealing temperature T (Figure 2). As the graph has shown, the critical nuclei radius increases as the particle size decreases, yet for nanoparticles with radii larger than 15nm, such dependence is trivial. Also, $r_c$ increases with the increase of annealing temperature, since higher temperature makes the $L1_0$ phase less thermodynamically stable. At the upper left corner of the graph denoting small particles annealed at high temperatures, $r_c$ even exceeds the value of 2R, meaning the formation of ordered domain under such conditions is impossible. This explains why small particles remain untransformed when annealed at high temperatures. It is also obvious that for nanoparticles, the size of the readily formed nuclei (ranging from 0.5 to 1nm) is significant comparing to the size of the particle.

To test the model's prediction on time-dependent properties of the system, we extracted data points from previous Monte-Carlo simulation studies[32] on the annealing of 5nm FePt nanoparticles at 1000 K and compared them with theoretical calculation result from Eq (12).

Unfortunately, no discrete experimental data concerning the domain expansion speed "g" is currently available. So, in order to validate the consistency of the model, we parameterized "g" using the first data point and obtained a trial value of 0.010 nm/s.

In Figure 3 the transformation curve for R=2.5nm T=1000 K is plotted alongside Monte-Carlo simulation data point (noted by open squares). Even though "g" is parameterized using the first data point, the curve agrees well with the rest of the points and clearly bears the feature of a typical JMA function. It is crucial that we point out here; the part of the curve concerning long annealing time may deviate from actual scenarios. Since our model did not include a dynamic solution on the elimination of APBs, which will necessarily arrest domain growth as transformation continues to the late stage, the curve is only accurate for low values of t.

Using Eq (12), it is possible to calculate the exact R-t function for any given transformed volume fraction X at annealing temperature T. In order to investigate the size dependence of the annealing time, we calculated the time needed to obtain 50% transformation (X=0.5) in particles with different sizes (Figure 4). In Figure 4 the solid blue and dashed red line represent the theoretical predictions for annealing temperature of 1000K and 770K respectively. It is found that, the annealing time required for ordered domain inside these nanoparticles to reach certain volume fraction decreases along with the size of the particle. This is again the result of surface nucleation and increased surface/volume ratio in smaller particles. As particle size increases, the amount of surface nucleation site per unit volume decreases, making the transformation of larger particles more time consuming. The blue circle represents the computer simulation result. The slight misplacement may either be caused by the inherent deviance of the model from real situations introduced by the approximation $dX = (1-X)dX_e$, the data extraction error, or the accumulated inaccuracy of the simulation.

To the best of our knowledge, no experimental or simulation data is available for the determination of expansion speed "g" at 770 K, so we used the corresponding value of 1000 K to compute a trial curve (dashed red line in Figure 4). Delalande *et al.* reported[35] that heating a nanoparticle with 2nm radius at 670 K for 24 min and subsequently at 770 K for 1 min yielded a single ordered domain roughly 50% of the particle size. Because the particle has been pretreated for an extensive period of time at 673 K, when annealing at 770 K it probably had readily formed nuclei. To take this into account, we subtracted the average time need to form a surface nuclei at 770 K (39.7 s) from the previous calculation, and plotted the result using solid red line. Surprisingly, the experiment data fit well with the calculation result despite the fact that we adopted "g" corresponding to 1000 K instead of 770 K. It might be that the smaller nuclei formed at 670 K require extra time to transform before they can be used as growing platforms at 770K. Yet the deviation caused by adopting a higher value of "g" cancels with this extra time, reducing the total amount of deviation.

The annealing process is quite swift as demonstrated above (half of the volume transformed in less than 100 s). But in experiments, it often requires thousands of seconds of annealing to obtain ordered particles. This shows the major barrier in the ordering process do not come from nucleation or domain growth. Instead, it might have to do with the elimination of APBs. Because the initial ordering directions of the nuclei are random, as ordered domains grow, APBs form between different domains and the growth ceases, causing the boundary atoms to remain disordered. These defects in nanoparticles can be difficult to remove. Since their elimination requires destruction of established ordered structure. Even though single crystal particles should be thermodynamically more stable, the activation energy for the Ostwald ripening like process[43] to produce them may be high enough to require additional annealing time to overcome.

Such an argument is well supported by the in situ experiment conducted by Delalande *et al.* In one of their experiment, nanoparticles with radius of 2 nm were kept at 820 K for 7 min and the TEM image showed partial transformation starting from the surface, agreeing with our model's surface nucleation assumption. However the state of the particle remained unchanged from 7 to 35 min, suggesting the thermal dynamic driving force was unable to overcome the kinetic barrier to facilitate the growth of ordered domains, thus keeping the particle at a metastable state. Elevate the temperature to 870 K, and the particle nearly completed its transformation in 5 min. At this stage, TEM images recorded 3 separate domains divided by APBs. A further 20 min annealing at 920 K was required to complete the elimination of these defects and produce an entirely $L1_0$-ordered single-crystalline nanoparticle, agreeing with our conclusion that the elimination of APBs is the most time consuming process.

Interestingly, these results pointed out a possible way of improving the annealing technique for FePt nanoparticles. If somehow an external factor can be introduced to give nuclei with certain ordering orientation extra advantages in formation or domain coarsening, it might prove to be a superior way to directly obtain good quality single crystalline nanoparticles than APB elimination through thermally activated atomic diffusion. Already, some preliminary research[47] has shown that external magnetic field can play a constructive role in the order-disorder transition of FePt nanoparticles. The exact mechanisms, however, is unfortunately not very well understood. The model we presented here offers a possible explanation.

The theoretical prediction of our model, however, does not mean smaller particles are easier to transform into $L1_0$ phases. As pointed out by Qi *et al*,[36] $L1_0$ phases in smaller nanoparticle are less thermodynamically stable and thus require lower annealing temperatures to produce. Yet the entire kinetic process is controlled by atomic diffusion, which slows exponentially with the decrease of temperature. The dilemma poses substantial difficulty in the ordering of small nanoparticles, even though under the same temperature they are kinetically more active.

There have been numerous reports[27] on the ordering size limit of $L1_0$ FePt nanoparticles. Conventionally, a particle diameter larger than 3 nm is thought to be necessary to induce order-disorder transformation. However, the existence of L1$_0$ ordered ultra-small nanoparticles (<3 nm) is nonetheless confirmed by several experiments. Based on our model, nanoparticles are theoretically transformable as long as their diameters are larger than the critical nuclei radius. This makes the ordering size-limit temperature dependent (see Figure 2), and offered a good explanation on previous contradictory reports.

Based on Eq (8), Eq (9) and the condition $r_c=2R$, we computed a curve representing the threshold particle diameter for ordering to occur (Figure 5). Particles annealed on the right side of the curve are transformable in theory while those annealed on the left side are not. Clearly, all the ultra-small FePt nanoparticles' reported transformability,[20, 35, 48-50] plotted as data points in the graph, agreed with the model's prediction. Particularly, the annealing results for 2 nm nanoparticles obtained by Miyazaki et al.[27] and Nandwana et al.[50] gave a good sense on the accuracy of the model's prediction. At the temperature of 973 K, nanoparticles with size smaller than 2 nm remain disordered regardless of the annealing time. Yet, just 50 K lower, crossing into the curve's right side, Nandwana et al. was able to obtain completely ordered 2 nm FePt nanoparticles through annealing. Moreover, these ordered 2 nm nanoparticles are present in the form of single crystals, suggesting the critical nuclei radii are close to the diameters of the particles, further lending credence to the model's validity.

In our previous thermodynamic studies[36], to address the influence of particle morphology on order-disorder transition, we introduced a shape factor "α" into our equation as a convenient way to account for the additional surface areas created by the presence of polyhedral shapes. The factor "α" is defined as the surface area ratio between non-spherical and spherical nanoparticles of identical volumes. For example, a nanoparticle with tetrahedral shape would have an α of 1.49 and a spherical particle would have an α of 1. So, for polyhedral nanoparticles, Eq (8) becomes:

$$T_c = (1 - \alpha \cdot \frac{1.691 r_0}{R}) T_{c,b} \qquad (14)$$

To assess how the shape factor influences the critical nuclei radius, which is an important indicator on the feasibility of A1-L1$_0$ transition, we plotted their relation using Eq (9) and Eq (14) with respect to several typical parameter settings. As shown in Figure 6, the critical nuclei radius increases with the increase of α. But such effects are trivial, especially at low temperatures or in particles with radius larger than 2 nm. Only for particles with radii of 1 nm, the deviance in shape factors marked a qualitative difference, under the annealing temperature of 850 K, particles with α<1.28 are transformable while particles with α>1.28 are not. Most FePt nanoparticles reported in literature have R larger than 1 nm and α smaller than 1.2 (truncated octahedron). Therefore, their ordering should not deviate much from those of spherical particles. For sure, the exact influence of morphology on transition kinetics is much more complicated. Both matrix composition and crystallographic orientation cast influences on the detailed energy structure of nanoparticles. However, here, we only intend to develop a simple model for general use. More demanding solutions should be obtained by case-sensitive numerical computation instead.

**IV. Model versatility and limitations**

In principle, the model can be used to quantitatively explain any interface-controlled solid state order-disorder transitions in nanoscale systems, provided that both size-dependence of transition temperature and the surface nucleation rate can be accurately formulated. Moreover, if the domain expansion speed can also be formulated against temperature, the model can be used to compute 3D transition surface as a function of temperature and time, X = f(T,t). However, if the model is to be applied to other systems, following limitations and prerequisites should be noted:

(1) For transition accompanied by surface segregation, the free-surface energy difference caused by the segregation needs to be accounted for in Eq (5). That is, θ can no longer be approximated by π/2.

(2) Elastic strain energy $\triangle G_e$ may not be negligible if the transition invokes relatively large lattice deformation. Importantly, for transitions with ε>5%, the new phase takes on a disk-like shape and the model is no longer applicable.

(3) For diffusion controlled transitions, the model only offers limited applicability. Because the

involvement of long-range transportation of atoms will inevitably make the domain expansion speed and nucleation energy barrier time-dependent, something the model we proposed here is not designed to coupe with.

## V. Conclusion

Based on the discussion above, it is clear that order-disorder transition kinetics in FePt nanoparticles has a strong dependence on the spacial dimension of the particles. With the increase of surface to volume ratio, nanoparticles become kinetically more active; yet the transition become thermodynamically less favorable. There exists a size limit in ordering. The fact that such limit is constrained by both kinetic nucleation barrier and atomic diffusion activation energy explains previous contradictory reports. The method proposed here should be able to be extended to apply to other alloy systems that experiences order-disorder transition, although this remains to be tested by future works.

**Acknowledgements:**

This work was supported by Distinguished Young Scholars Foundation of Hunan Province (No. 13JJ1002) and Shenghua Scholar Program of Central South University.

# Reference

[1]  A. Moser, K. Takano, D. T. Margulies, M. Albrecht, Y. Sonobe, Y. Ikeda, S. H. Sun, and E. E. Fullerton, J Phys D Appl Phys **35**,R157, (2002).

[2]  S. H. Sun, C. B. Murray, D. Weller, L. Folks, and A. Moser, Science **287**,1989, (2000).

[3]  J.-Y. Bigot, H. Kesserwan, V. Halte, O. Ersen, M. S. Moldovan, T. H. Kim, J.-t. Jang, and J. Cheon, Nano Letters **12**,1189, (2012).

[4]  C. P. Luo, and D. J. Sellmyer, Appl Phys Lett **75**,3162, (1999).

[5]  C.-b. Rong, D. Li, V. Nandwana, N. Poudyal, Y. Ding, Z. L. Wang, H. Zeng, and J. P. Liu, Advanced Materials **18**,2984, (2006).

[6]  B. Stahl, N. S. Gajbhiye, G. Wilde, D. Kramer, J. Ellrich, M. Ghafari, H. Hahn, H. Gleiter, J. Weissmuller, R. Wurschum, and P. Schlossmacher, Advanced Materials **14**,24, (2002).

[7]  O. Dmitrieva, B. Rellinghaus, J. Kastner, and G. Dumpich, J Cryst Growth **303**,645, (2007).

[8]  T. Burkert, O. Eriksson, S. I. Simak, A. V. Ruban, B. Sanyal, L. Nordström, and J. M. Wills, Phys Rev B **71**,134411, (2005).

[9]  Y. Kota, and A. Sakuma, J Appl Phys **111**,07A310, (2012).

[10] R. V. Chepulskii, and W. H. Butler, Appl Phys Lett **100**, 142405,142405, (2012).

[11] C. Antoniak, J. Lindner, M. Spasova, D. Sudfeld, M. Acet, M. Farle, K. Fauth, U. Wiedwald, H. G. Boyen, P. Ziemann, F. Wilhelm, A. Rogalev, and S. H. Sun, Physical Review Letters **97**,117201, (2006).

[12] H. G. Boyen, K. Fauth, B. Stahl, P. Ziemann, G. Kastle, F. Weigl, F. Banhart, M. Hessler, G. Schutz, N. S. Gajbhiye, J. Ellrich, H. Hahn, M. Buttner, M. G. Garnier, and P. Oelhafen, Advanced

Materials **17**,574, (2005).

[13] B. Rellinghaus, S. Stappert, M. Acet, and E. F. Wassermann, J Magn Magn Mater **266**,142, (2003).

[14] K. Sato, B. Bian, T. Hanada, and Y. Hirotsu, Scripta Mater **44**,1389, (2001).

[15] Y. Tsuji, S. Noda, and S. Nakamura, J Vac Sci Technol B **29**,031801, (2011).

[16] M. L. Plumer, J. Van Ek, and D. Weller, *The physics of ultra-high-density magnetic recording* (Springer, Berlin ; New York, 2001), Springer series in surface sciences,, 41.

[17] V. Monnier, M. Delalande, P. Bayle-Guillemaud, Y. Samson, and P. Reiss, Small **4**,1139, (2008).

[18] M. Nakaya, Y. Tsuchiya, K. Ito, Y. Oumi, T. Sano, and T. Teranishi, Chem Lett **33**,130, (2004).

[19] S. H. Sun, Advanced Materials **18**,393, (2006).

[20] K. E. Elkins, T. S. Vedantam, J. P. Liu, H. Zeng, S. H. Sun, Y. Ding, and Z. L. Wang, Nano Letters **3**,1647, (2003).

[21] T. Herricks, J. Y. Chen, and Y. N. Xia, Nano Letters **4**,2367, (2004).

[22] J. Kim, C. Rong, J. P. Liu, and S. Sun, Advanced Materials **21**,906, (2009).

[23] J. Sort, S. Surinach, M. D. Baro, D. Muraviev, G. I. Dzhardimalieva, N. D. Golubeva, S. I. Pomogailo, A. D. Pomogailo, W. A. A. Macedo, D. Weller, V. Skumryev, and J. Nogues, Advanced Materials **18**,466, (2006).

[24] Q. Yan, A. Purkayastha, T. Kim, R. Kroeger, A. Bose, and G. Ramanath, Advanced Materials **18**,2569, (2006).

[25] Z. R. Dai, S. H. Sun, and Z. L. Wang, Nano Letters **1**,443, (2001).

[26] R. V. Chepulskii, and W. H. Butler, Phys Rev B **72**, 134205,134205, (2005).

[27] T. Miyazaki, O. Kitakami, S. Okamoto, Y. Shimada, Z. Akase, Y. Murakami, D. Shindo, Y. K. Takahashi, and K. Hono, Phys Rev B **72**,144419, (2005).

[28] M. E. Gruner, J Phys D Appl Phys **43**,474008, (2010).

[29] A. Alam, B. Kraczek, and D. D. Johnson, Phys Rev B **82**,024435, (2010).

[30] M. Imaizumi, C. A. Soufen, C. A. F. Pintao, L. C. Varanda, and M. Jafeficci, Mat Sci Eng a-Struct **521-22**,167, (2009).

[31] M. Muller, and K. Albe, Phys Rev B **72**,094203, (2005).

[32] M. Muller, and K. Albe, Beilstein J Nanotech **2**,40, (2011).

[33] Z. R. Dai, S. H. Sun, and Z. L. Wang, Surf Sci **505**,325, (2002).

[34] F. Tournus, K. Sato, T. Epicier, T. J. Konno, and V. Dupuis, Physical Review Letters **110**, 055501,055501, (2013).

[35] M. Delalande, M. J. F. Guinel, L. F. Allard, A. Delattre, R. Le Bris, Y. Samson, P. Bayle-Guillemaud, and P. Reiss, J Phys Chem C **116**,6866, (2012).

[36] W. H. Qi, Y. J. Li, S. Y. Xiong, and S. T. Lee, Small **6**,1996, (2010).

[37] D. A. Porter, and K. E. Easterling, *Phase transformations in metals and alloys* (Chapman & Hall, London ; New York, 1992), 2nd edn.

[38] X. Li, Z. H. Li, X. Liu, Y. B. Li, J. M. Bai, F. L. Wei, and D. Wei, Ieee T Magn **46**,2024, (2010).

[39] Y. Chen, S. Iwata, and T. Mohri, Calphad **26**,583, (2002).

[40] N. A. Schclar, *Anisotropic analysis using boundary elements* (Computational Mechanics Publications, Southampton, UK ; Boston, 1994), Topics in engineering,, 20.

[41] J. S. Kim, Y. M. Koo, B. J. Lee, and S. R. Lee, J Appl Phys **99**,053906, (2006).

[42] A. Dannenberg, M. E. Gruner, A. Hucht, and P. Entel, Phys Rev B **80**,245438, (2009).

[43] J. Schmelzer, *Nucleation theory and applications* (Wiley-VCH, Weinheim, 2004).

[44] K. A. Jackson, *Kinetic processes : crystal growth, diffusion, and phase transformations in materials* (Wiley-VCH; John Wiley distributor, Weinheim; Chichester, 2010), 2nd completely rev. and enl. edn.

[45] T. J. Klemmer, N. Shukla, C. Liu, X. W. Wu, E. B. Svedberg, O. Mryasov, R. W. Chantrell, D. Weller, M. Tanase, and D. E. Laughlin, Appl Phys Lett **81**,2220, (2002).

[46] X. H. Li, B. T. Liu, W. Li, H. Y. Sun, D. Q. Wu, and X. Y. Zhang, J Appl Phys **101**,093911, (2007).

[47] H. Y. Wang, X. K. Ma, Y. J. He, S. Mitani, and M. Motokawa, Appl Phys Lett **85**,2304, (2004).

[48] R. Medwal, N. Sehdev, and S. Annapoorni, J Phys D Appl Phys **45**,055001, (2012).

[49] L. E. M. Howard, H. L. Nguyen, S. R. Giblin, B. K. Tanner, I. Terry, A. K. Hughes, and J. S. O. Evans, Journal of the American Chemical Society **127**,10140, (2005).

[50] V. Nandwana, K. E. Elkins, N. Poudyal, G. S. Chaubey, K. Yano, and J. P. Liu, J Phys Chem C **111**,4185, (2007).

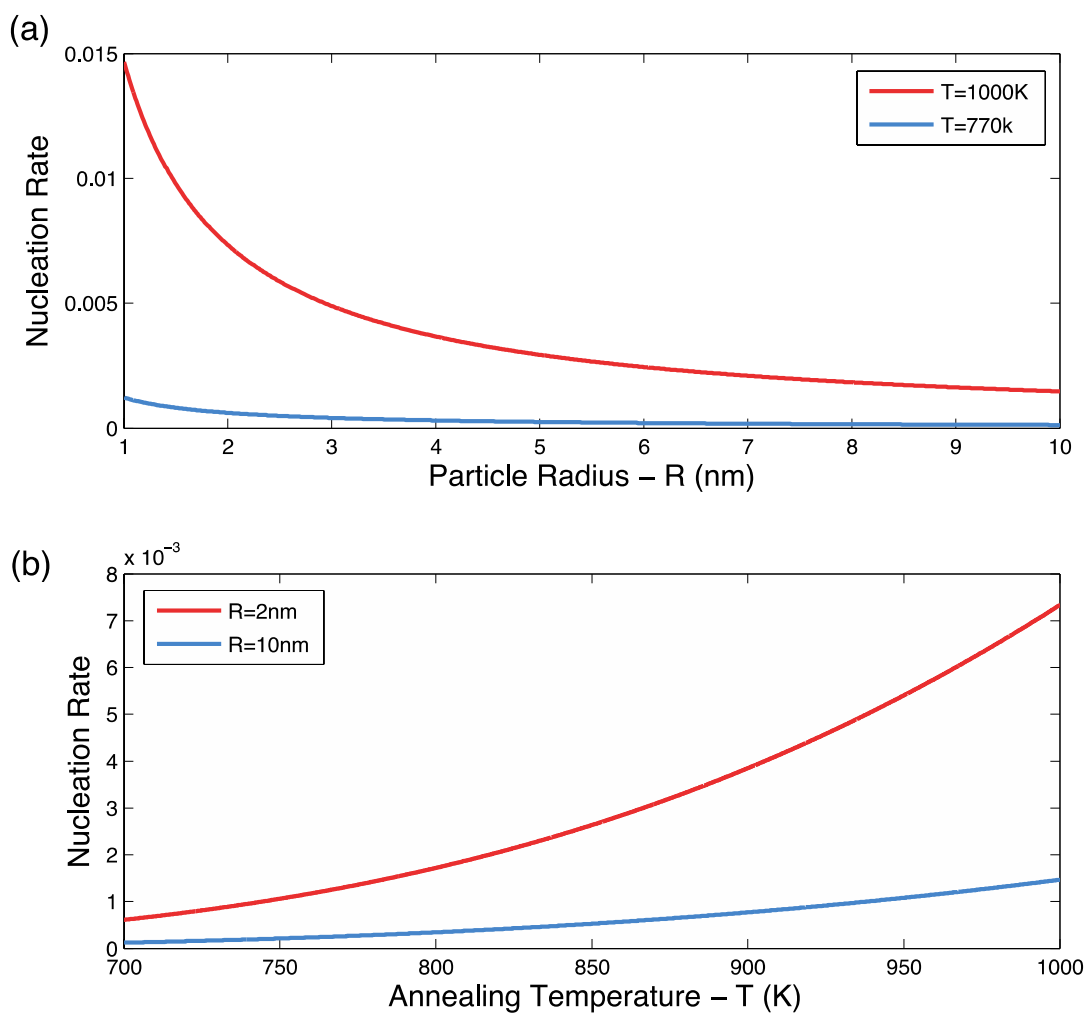

Figure 1

**Figure 1** (a) The nucleation rate as a function of nanoparticle radius. The red line corresponds to particles annealed at 1000 K, the blue, 770 K. (b) The nucleation rate as a function of annealing temperature. The red line corresponds to particles with radius of 2 nm, the blue, 10 nm.

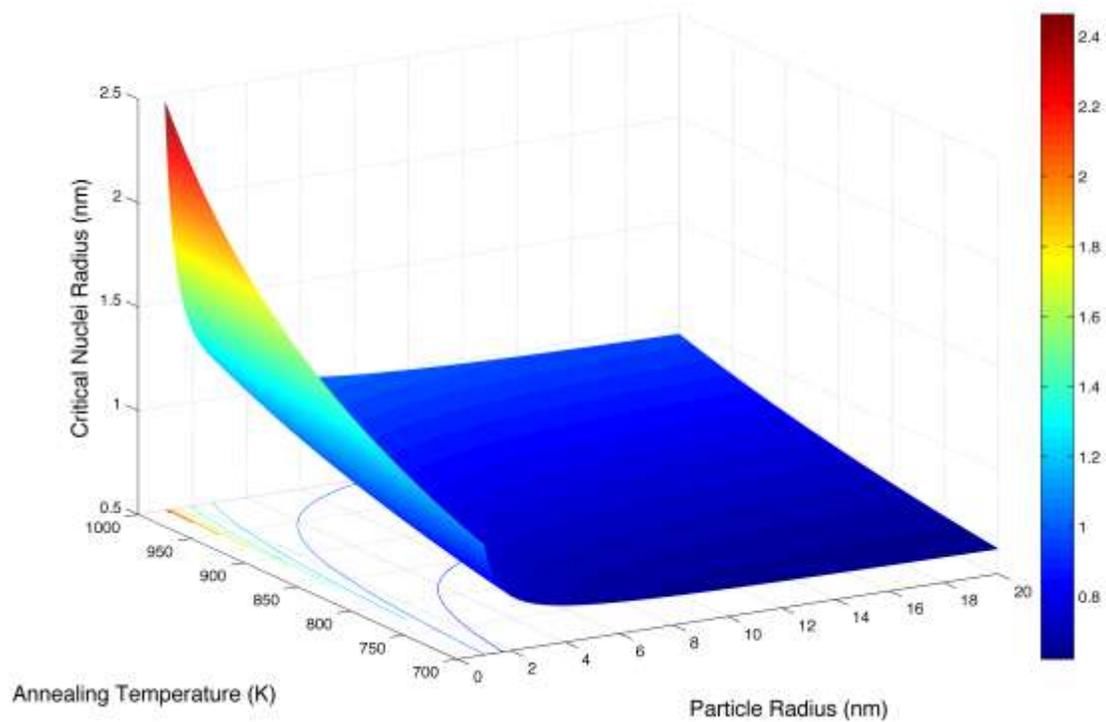

Figure 2

**Figure 2** Critical nuclei radius as a function of particle radius (from 1 to 20 nm) and annealing temperature (from 700 to 1000 K). Contour lines are projected on the base plane to quantitatively illustrate the 3-D features of the graph.

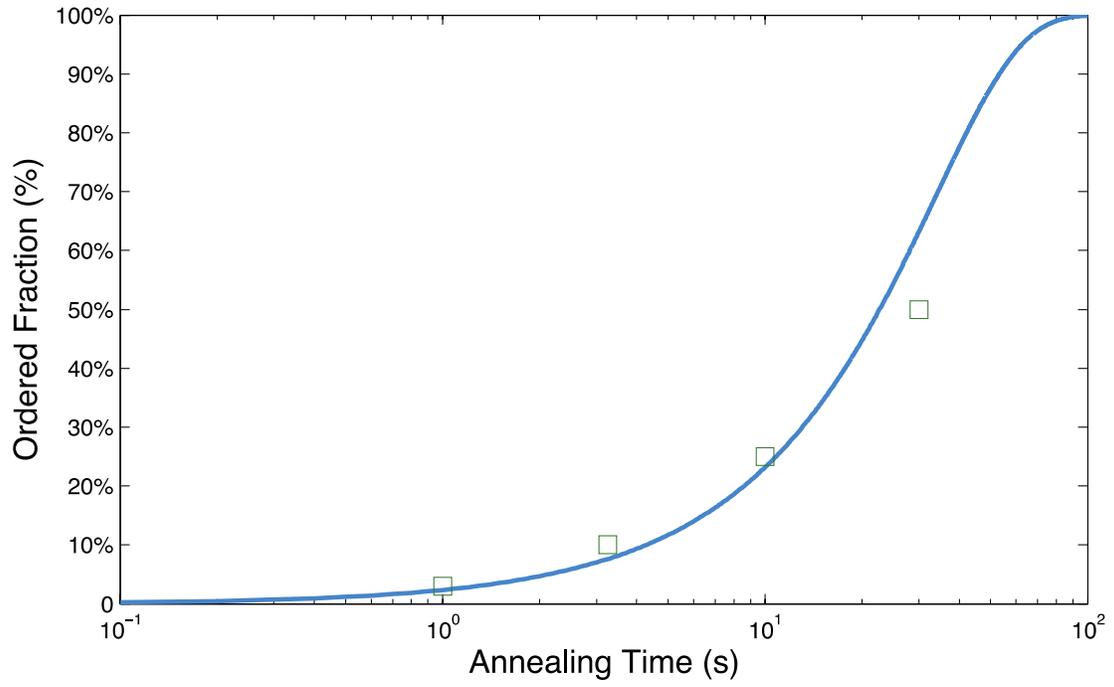

Figure 3

**Figure 3** Transformed volume fractions as a function of annealing time. The open squares represent the Monte-Carlo simulation result.

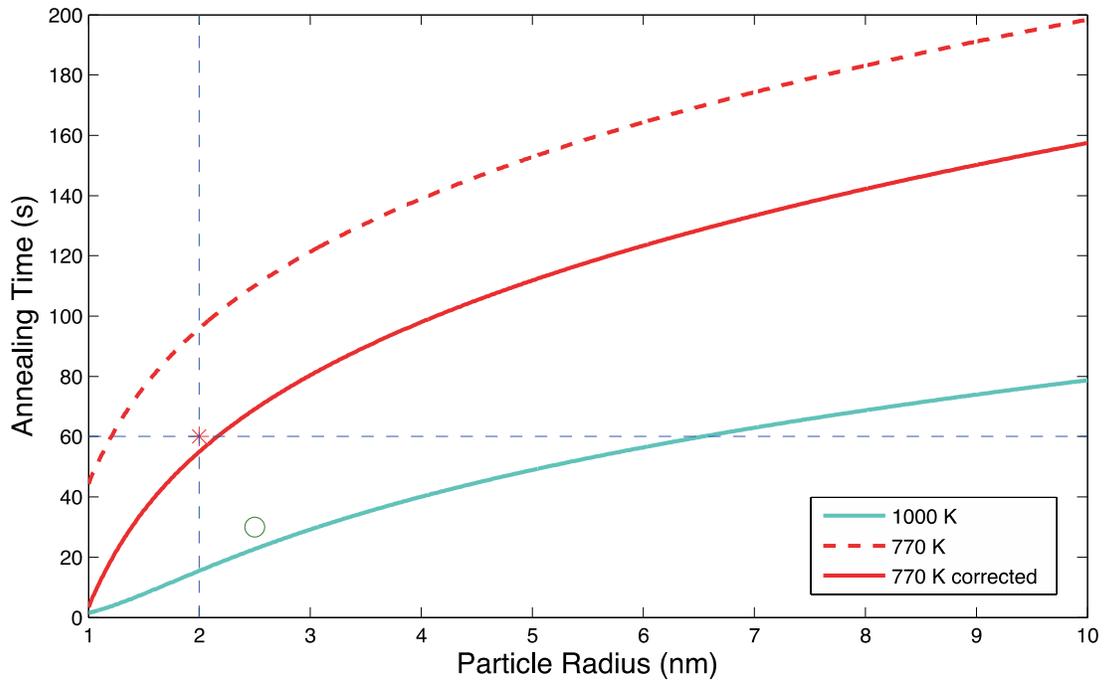

Figure 4

**Figure 4** Annealing time required to transform 50% of the particle volume as a function of particle radius. The solid blue line denotes the theoretical calculation result for annealing temperature of 1000 K while the dashed red line denotes the calculation results for 770 K. The solid red line represents the calculation result for 770 K after nucleation time correction. The blue circle and red cross represent the simulated value and experiment result respectively.

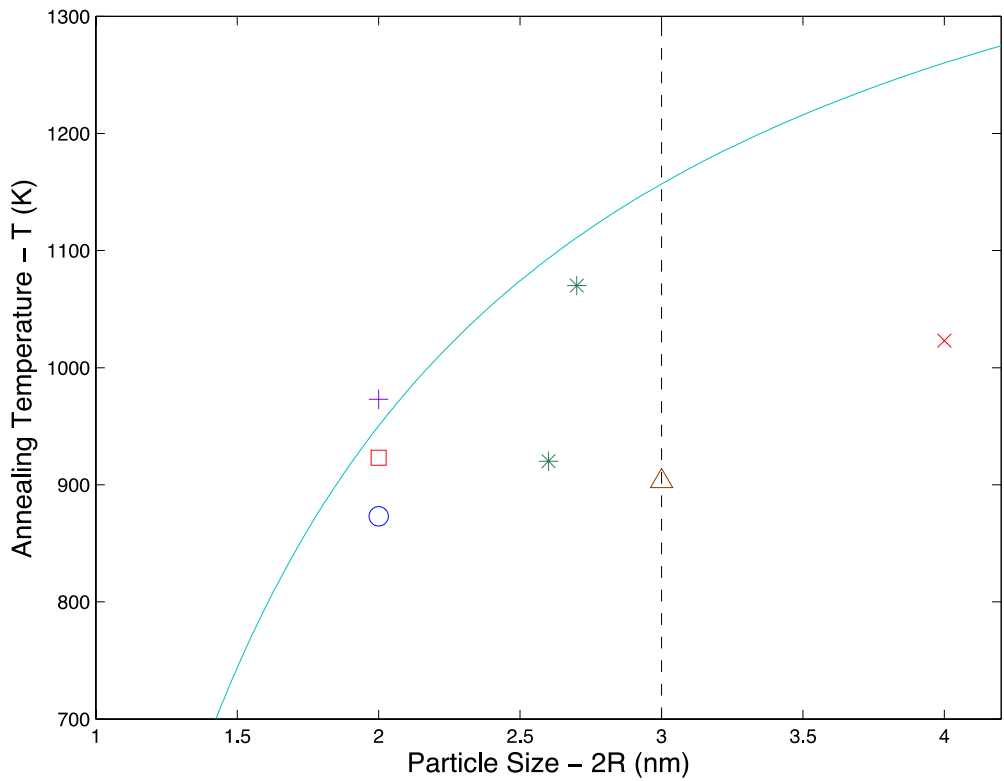

Figure 5

**Figure 5** Size limit for order-disorder transition as a function of annealing temperature. The red cross, brown triangle, green asterisks, blue circle, red square represent the experimental observations of $L1_0$ ordered nanoparticles reported in Ref 46, 47, 34, 19, 48 respectively. The purple plus sign represents the experimental observation[27] of untransformed A1 disordered nanoparticles after high temperature annealing.

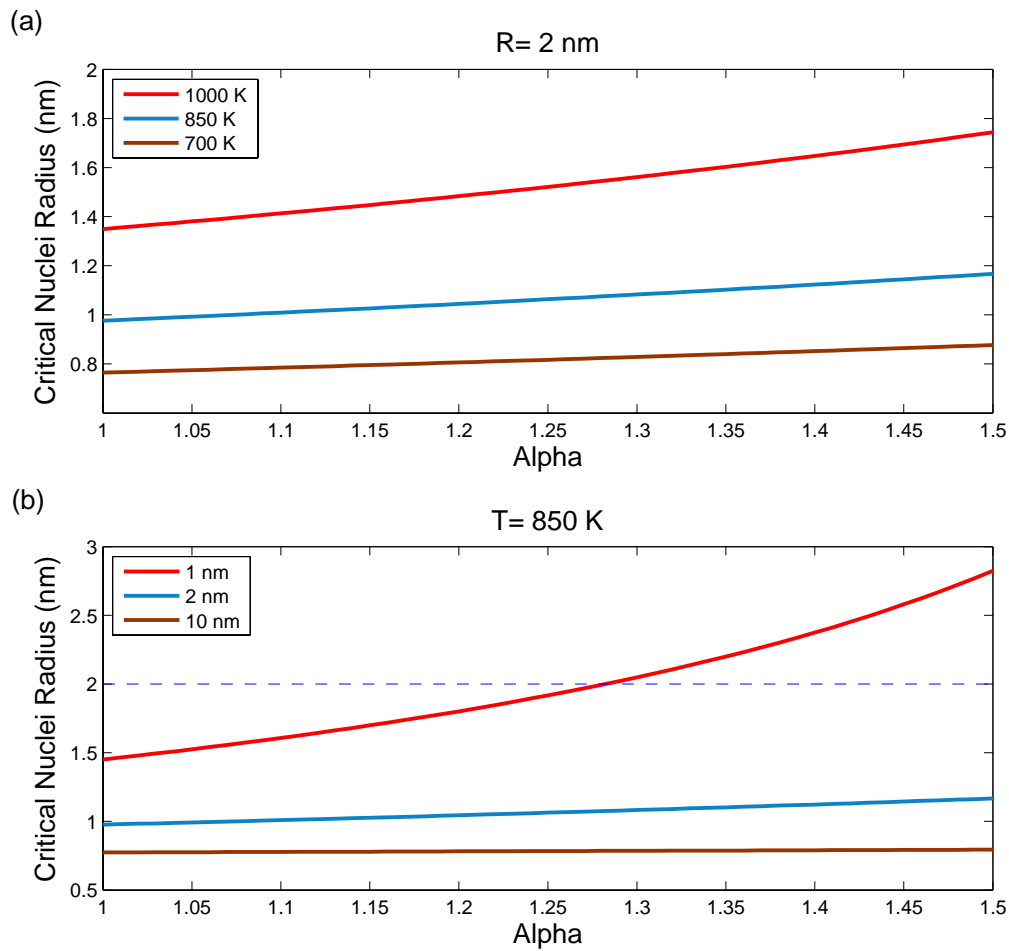

Figure 6

**Figure 6** (a) For nanoparticles with radius of 2 nm, critical nuclei radius as a function of shape factor α. The red line corresponds to particles annealed at 1000 K, the blue 850 K, the brown 700 K. (b) At the annealing temperature of 850 K, critical nuclei radius as a function of shape factor α. The red line corresponds to particles with radius of 1 nm, the blue 2 nm, the brown 10 nm.